# Waveguide-integrated electroluminescent carbon nanotubes

Svetlana Khasminskaya[1]*, Feliks Pyatkov[1,2,3]*, Benjamin S. Flavel[1], Wolfram H. Pernice[1,2], Ralph Krupke[1,2,3]

[1] Institute of Nanotechnology, Karlsruhe Institute of Technology, 76021 Karlsruhe, Germany

[2] DFG Center for Functional Nanostructures (CFN), 76031 Karlsruhe, Germany

[3] Department of Materials and Earth Sciences, Technische Universität Darmstadt, 64287 Darmstadt, Germany

* The authors contributed equally to this work.

Corresponding authors: R. Krupke (krupke@kit.edu), W.H. Pernice (wolfram.pernice@kit.edu)

**Carbon based optoelectronic devices promise to revolutionize modern integrated circuits by combining outstanding electrical and optical properties into a unified technology [1,2]. By coupling nanoelectronic devices to nanophotonic structures functional components such as nanoscale light emitting diodes [3], narrow-band thermal emitters [4], cavity controlled detectors [4,5] and wideband electro-optic modulators [6] can be realized for chipscale information processing. These devices not only allow the light-matter interaction of low-dimensional systems to be studied, but also provide fundamental building blocks for high bandwidth on-chip communication [7,8]. Here we demonstrate how light from an electrically-driven carbon-nanotube can be coupled directly into a photonic waveguide architecture. We realize wafer-scale, broadband sources integrated with nanophotonic circuits allowing for propagation of light over centimeter distances. Moreover, we show that the spectral properties of the emitter can be controlled directly on chip with passive devices using Mach-Zehnder interferometers and grating structures. The direct, near-field coupling of electrically generated light into a waveguide, opposed to far-field fiber coupling of external light sources, opens new avenues for compact optoelectronic systems in a CMOS compatible framework.**

Photons propagating at the speed of light and outpacing electrons are the fastest carriers of information possible. For this reason a large proportion of mid- and long-distance electrical communication connections have been replaced by fiber optics in the last years [9]. The next evolutionary step will be the replacement of short-distance electrical connections by optical waveguides, which will thereby enhance on-chip data transfer rates, for example between processor and memory [10]. In order to achieve this goal the development of optical modules with large numbers of input and output channels will be required. Furthermore, on-chip generation of light will be needed to overcome foreseeable limitations in scalability and reproducibility. The then required level of integration will exceed the capacity of conventional integrated optical circuits [11] and will necessitate the use of waveguides with tight modal confinement, as well as the co-integration of active components.

Since the beginning of the field it has been a challenge to couple light into nanoscale photonic waveguides. The current state-of-the-art solution is to launch light from external sources, such as lateral microcavity lasers [12], vertical cavity surface emitting lasers [13], or microdisc lasers [14], into the photonic waveguides using fiber-coupling techniques [15,16]. These hybrid solutions, however, require sophisticated multilevel nanofabrication processes for producing the lasers, which is in marked contrast to comparably simple photonic waveguide processing techniques. Furthermore, tight alignment tolerances for the orientation of the waveguide with respect to an optical fiber core make such an approach unfeasible for large numbers of input/output ports.

An alternative approach is the integration of monolithic nanoscale light emitters with photonic waveguide structures. A first proof-of-principle has been shown by Park et al., who placed a nanowire light source close to a photonic crystal waveguide [17]. More recently also carbon nanotubes have been combined with photonic waveguides and cavities [18,19]. In these experiments the nanotube light emission has been stimulated optically by a free-space laser source and coupling of fluorescent light from nanotubes into photonic structures has been demonstrated. Carbon nanotubes, however, can also be stimulated electrically to emit light, which is the preferred method of excitation for chipscale solutions. Depending on the type of the nanotube, its diameter and the mode of operation the emission can be narrowband or broadband [20,21,22,23,24]. Moreover the emitted light is polarized along the nanotube axis and emitted preferentially perpendicular to the nanotube axis [20]. In this work we demonstrate efficient coupling of light emitted from an electrically-stimulated carbon nanotube into a photonic waveguide structure by integrating a carbon nanotube with its long axis perpendicular to a photonic waveguide. The work is based on electric-field assisted placement of solution-processed carbon nanotubes into pre-patterned structures containing the photonic waveguide, couplers, Mach-Zehnder structures and electrical wiring. Our approach allows for contacting multiple devices in parallel as a key step towards carbon based optical interconnects.

**Integration of carbon nanotube emitters with nanophotonic circuits**

All the devices consist of three components: carbon nanotubes, nanophotonic waveguides with coupler gratings, and metallic contacts. The device fabrication was performed in three steps. At first, 60 nm thick metal contacts with a gap of 1 µm were formed on a doped Si-wafer with $SiO_2$ (2 µm) / $Si_3N_4$ (0.2 µm) top layers, using electron-beam lithography and subsequent metal evaporation. Then, 500 nm wide waveguides terminated with focusing grating couplers were defined with electron beam lithography and formed by thinning 100 nm of the $Si_3N_4$ layer in between the metal contacts by reactive ion etching. Optimal etching parameters for obtaining the nominal etching depth in combination with minimal surface roughness were identified using reflectometry, scanning electron microscopy and atomic force microscopy. Finally, single-walled carbon nanotubes were deposited in between the metal contacts and onto the waveguide by dielectrophoresis from an aqueous dispersion [25,26]. The dielectrophoretic force thereby ensures precise alignment of the nanotubes with the nanotube axis perpendicular to the direction of the waveguide. The nanotube dispersion contained predominantly semiconducting nanotubes and residual metallic nanotubes. Variation of the dielectrophoresis parameters like exposure time, concentration of nanotubes as well as amplitude and frequency of the applied voltage was used to control the number of carbon nanotubes deposited in between the metal contacts. For details on the device fabrication and materials we refer to the methods section.

Fig. 1a shows the central device structure comprising of a nanotube which is positioned across the waveguide and is in contact with the metal electrodes. We will refer to this structure as the waveguide-coupled carbon nanotube emitter E in the following. A schematic cross-section of E is shown in Fig. 1b. The chosen structure enables single-mode propagation of visible and near-infrared light within the waveguide, and ensures that light propagation is not perturbed by the 200 nm distant metal electrodes. At the same time a direct contact between the nanotube sidewall and the waveguide top surface, and between the nanotube end and metal electrodes, respectively, is ensured. In all devices, the waveguides are terminated by two Bragg couplers C1 and C2, as shown in Fig. 1c. This configuration allows the far-field coupling of light into and out of the waveguide for transmission experiments using an external light source as described in the methods section. The grating couplers provide typical insertion loss of 7dB at 700-800 nm wavelength if the $Si_3N_4$ layer is thinned to half of its thickness, as shown in Fig. 1b. The coupling bandwidth is roughly 30 nm, centered around a coupling wavelength that can be adjusted by varying the period of the grating (supplementary information). The couplers C1 and C2 also ensure that light which is generated by the carbon nanotube emitter E - coupled into the waveguide and propagating along the waveguide - will be coupled out for far-field detection. The expected near-field coupling of light from the nanotube into the waveguide has been simulated with 3D finite-difference time-domain calculations of the electric-field intensity $\vec{E}^2$, described in the methods section. The nanotube emitter is modeled as an electrical dipole at the waveguide-air interface with the dipole moment oscillating perpendicular to the waveguide axis and hence parallel to the nanotube axis. Fig. 1d shows, that near-field coupling of light from the nanotube emitter into the waveguide is expected, as well as the propagation of light within the waveguide. A cross-section through the waveguide mode at 750 nm is shown in Fig. 1b as a color overlay on the waveguide facet. Fig. 1d also shows that part of the emitted light will be far-field coupled into and out of the substrate along the surface normal. These light paths give rise to interference effects shaping the spectral profile of the emitter and allow for free-space spectral analysis of the emitted light.

**Electroluminescence from waveguide-coupled carbon nanotubes**

Electroluminescence from our devices has been measured in the dark with a low-temperature CCD camera directly attached to an optical microscope. The electroluminescence images were then compared with recordings under external illumination, enabling identification of light emission sources. Spectral information was obtained through a retractable grating in combination with a field aperture. The device was mounted on a motorized stage and electrically connected through probe needles. For details on the electroluminescence measurement setup we refer to the methods section. Upon electrical biasing of the nanotube in the range of typically 3-10 V (corresponding to an electrical power of 1-100 µW) we observe light emission from three spots as shown in Fig. 2a. The central emission spot is correlated with the position of the waveguide-coupled carbon nanotube emitter E and hence due to the emission of photons along the surface normal within 60° acceptance angle of the microscope objective. The other two emission spots are located at the positions of the coupler gratings C1 and C2. Light is generated at E and the photons detected at C1 and C2 have propagated through the waveguide prior to being diffracted at the grating coupler. The image is thus direct evidence, that photons emitted from an electrically-driven carbon-nanotube, couple into the waveguide, propagate over 150 µm in both directions of the waveguide, and couple out into the far field. Comparing the peak signal intensities recorded at E, C1 and C2, which are similar within one order of magnitude, we find that the near-field coupling is rather efficient, in particular when taking into account the insertion loss of the grating couplers. Fig. 2b shows a series of emission spectra recorded at E with increasing electrical power. Overall the signal intensity increases towards longer wavelength with a wavelength-dependent modulation. The spectra $I(\lambda)$, measured at the detector, show the typical profile of black-body radiation enveloped with the interference fringes due to back-reflection from the underlying silicon substrate. Photons emitted into air ($I_1(\lambda)$) are interfering with photons emitted towards the silicon substrate and reflected back at the SiO$_2$/Si interface ($I_2(\lambda)$), as described by $I(\lambda) = I_1(\lambda) + I_2(\lambda) + 2 \cdot \sqrt{I_1(\lambda) \cdot I_2(\lambda)} \cdot \text{Re}\,\gamma_{12}(\tau)$ [27]. $\gamma_{12}(\tau)$ is the complex degree of coherence function and $\tau$ is the delay time between the emitted and reflected photons. We have fitted the spectra to $f(\lambda) \propto \cdot f_{Planck}(\lambda,T) \cdot [1 + 2 \cdot \gamma_{12}(\tau) \cdot f_{int}(\lambda)]$, with the temperature $T$ dependent Planck curve $f_{Planck}(\lambda,T)$ and a structure dependent interference term $f_{int}(\lambda)$. The interference term has been calculated with a transfer matrix approach taking into account that the nanotube emitter is located at the Si$_3$N$_4$/air interface. For details on the fitting procedure we refer to the supplementary information.

In general the visibility of the interference fringes depends on the degree of temporal coherence and is determined by the magnitude of the coherence function $|\gamma_{12}(\tau)|$. A value of 1 is obtained for a fully coherent source and 0 for a completely incoherent source [27]. $\tau = \Delta z/c$ is defined by the optical path difference $\Delta z$. With $\gamma_{12}(\tau) = \int_0^\infty f_{Planck}(\nu) \cdot \exp(-2\pi i \nu \tau) d\nu$ [28], $\Delta z = 6.98$ μm [29], and using the extracted nanotube emitter temperatures T = 1395-1622K, we can calculate the theoretically expected value of $|\gamma_{12}(\tau)|$ for our thermal emitter and obtain $|\gamma_{12}(\tau)|$ = 0.0040-0.0026. Experimentally, however, we find a somewhat smaller value of $|\gamma_{12}(\tau)|$ = 0.025-0.016. This discrepancy could be related to the reduced dimension of the emitter - an argument that has been put forward to explain the enhanced coherence recently observed with metallic nanowires [30]. In our case we expect a long-wavelength cut-off because of the finite nanotube length, which would reduce the spectral width of the emitter and thus enhance the temporal coherence. However, studying these effects further is beyond the scope of this paper. That the emission spectrum does not reflect the narrow-band excitonic transitions of semiconducting nanotubes [20,21], but rather broad-band thermal emission [22,24] can be attributed to residual metallic nanotubes present in our devices, which would also explain the small on-off ratios of the transfer characteristics (not shown). Fabricating similar devices from ultra-high purity semiconducting nanotube dispersions would alleviate this effect.

We then study the spectral properties of the electroluminescent light that is coupled out by the coupler gratings C1 and C2. Fig. 2c is a high-resolution image of C2, superimposed on a scanning electron micrograph. The waveguide is entering the coupler grating from the top of the image. One can observe that the coupling of light out of the surface into the far-field occurs over an extended region. The maximum signal intensity is recorded at the position of the first grating lines (compare with supplementary information Fig. S1a) and is caused by diffusive scattering of photons with wavelengths outside the bandwidth of the grating coupler. Fig. 2d shows electroluminescence spectra recorded at the diffusive scattering region of C2 for two gratings with different period and thus different central coupling wavelengths. The overall spectra from the diffusive scattering regions are qualitatively similar to the emission spectra recorded at E, and the intensity modulations due to interference effects at the emitter E are also reproduced. However the spectra reveal distinct dips at the wavelengths which correspond to the characteristic grating wavelengths $\lambda_1$ and $\lambda_2$ defined by the grating period, respectively. Those "missing" photons continue to propagate into the grating structure and are then coupled out closer to the bottom of the coupler. Hence the spectra recorded at the Bragg scattering regions are dominated by photons with wavelength $\lambda_1$ and $\lambda_2$, respectively. As a result, grating structures inscribed into the on-chip photonic waveguide architecture may be employed for spectral filtering.

**Light propagation in extended waveguide structures**

For potential applications where electrical wiring is replaced by photonic leads it is important to demonstrate the coupling of light from an electroluminescent carbon nanotube into extended waveguides and to determine propagation losses within the waveguide. Therefore, we have fabricated a series of nanophotonic circuits with asymmetric long waveguides between the waveguide-coupled nanotube emitter E and the couplers C1 and C2. The distance $d_{E-C1}$ between E and C1 is fixed to 25 μm, whereas the distance $d_{E-C2}$ between E and C2 is varied systematically from 1.6 mm to 10.1 mm for the devices shown in Fig. 3a. Fabricating identical couplers within the field of view and keeping for all samples $d_{E-C1} \ll d_{E-C2}$, allows calculating the losses in the waveguide from the signal intensity at C1 and C2, even though the internal light source intensity varies from sample to sample. We first characterize the propagation loss in the photonic circuits by characterizing the transmission performance using a supercontinuum source and a spectrometer as shown in Fig.3b. The data reveals the Gaussian profile of the grating couplers, where the interference fringes result from Fabry-Perot-like reflections between the grating couplers. With increasing waveguide length higher absorption is present, leading to lower overall transmission within the spectrum. Then we characterize the on-chip losses using the emission from the CNT emitter. Fig. 3b shows the CCD-camera image of the electrically biased device with a waveguide of length L = 6.5 mm. Similar to Fig. 2a we observe light emission from E, C1 and C2. Also here, E is the only source of photons and hence the photons emitted at C2 are coupled out after propagation through the 6.5 mm long waveguide. The losses per unit length $\eta$ are calculated via $\eta = d/dL(10\log(I_{C1}/I_{C2}(L)))$. $I_{C1}$ and $I_{C2}$ are the signal intensities at C1 and C2 integrated within the indicated regions in Fig. 3b. Fig. 3d shows that the average loss per length is 9.1±2.1 dB/cm, which is typical for the type of waveguide geometry used here [31]. The results are also confirmed by the independent loss measurements with the external light source (methods section). The transmission spectra shown in Fig. 3c are dominated by the narrow-band couplers and therefore each spectrum has a maximum at 756 nm, corresponding to the characteristic wavelength of the used gratings. The overall signal intensity decreases with increasing waveguide length and is due to losses in the waveguide. Hence $\eta = d/dL \cdot \log(I_0 / I_{C2,765nm}(L))$, under the condition of identical fiber coupling for every sample. $I_0$ is the signal intensity of the external source determined with a reference detector. Fig. 3d shows that the average loss per length is 12.7±1.2 dB/cm and compares well to the measurement with the internal nanotube light source.

**On-chip interference experiments**

Finally, we realize Mach-Zender (MZ) interferometers co-integrated with a waveguide-coupled carbon nanotube emitter for the purpose of demonstrating electrically-driven interference of light on a chip. The MZ interferometer is located between the emitter E and the coupler C2 and consists of a bifurcated waveguide, that splits into two arms with a path difference $\Delta L = 50 \mu m$, before rejoining, as shown in Fig. 4a. The total waveguide length between C1 and C2 is 640 $\mu m$ when passing through the lower arm and 690 $\mu m$ for the longer arm, respectively. Biasing the emitter E induces light emission from E, C1 and C2, as observed in the previous structures. We focus now on the emission at the coupler C2 at the output of the MZ structure. Fig. 4b shows spatially resolved emission spectra of C2 along the y-axis. The dashed lines mark the location of C2 within the y-axis range. There are two types of wavelength-dependent intensity modulations: A long-wave modulation, with intensity minima at 800 nm and 900 nm, and a short-wave modulation with a period of a few nanometers. This can be better seen when integrating the spectra over the indicated y-axis range. The resulting data is plotted Fig. 4d together with a spectrum recorded at E. The long-wave intensity modulation reproduces the modulation directly measured at E, and hence is due to interference effects along the surface normal because of back-reflection from the silicon substrate. The short-wave intensity modulation, measured only at C2, is induced by the MZ structure and is hence due to the interference of light within the waveguide along the surface. The expected periodicity of the intensity modulation $\Delta \lambda$ (or free spectral range FSR), is given by $\Delta \lambda = \lambda^2 / (n_g(\lambda) \cdot \Delta L)$ [32], where $n_g(\lambda)$ is the group refractive index of the waveguide according to the supplementary information. With $n_g$ (850nm) = 1.964 and $\Delta L = 50 \mu m$ we obtain $\Delta \lambda = 7.4$ nm at 850 nm, which fits very well to the experimental value of ~ 7 nm. The data is confirmed by transmission measurements with an external light source, and similar short-wave intensity oscillation are observed, as shown in Fig. 4c. The periodicity of ~ 5 nm, measured at 745 nm, fits well to $\Delta \lambda = 5.4$ nm using $n_g$ (750nm) = 2.047. Fig. 4e shows the excellent agreement between FSR measurements obtained with the on-chip waveguide-coupled carbon nanotube emitter and the external light source, along with a theoretical simulation with no fit parameter.

**Conclusion**

The direct, near-field coupling of light from an electrically-driven carbon nanotube into a waveguide, as opposed to the traditional far-field fiber coupling of an external light source, opens up new opportunities to produce compact optoelectronic systems. Considering the wide range of different, structure-dependent emission spectra of semiconducting and metallic carbon nanotubes and the continuing progress in the sorting of specific nanotubes, it seems possible to use nanotubes with specific emission lines in the near future. The use of electrically triggered on-chip nanotube emitters for signal transmission through extended waveguides and interferometers shown in this work provides the basis for next-generation nanoscale interconnects that can be seamlessly integrated with passive silicon photonic technology.

**Methods**

**Device fabrication and materials**

Rib waveguides were fabricated in close vicinity to electrical contacts using several steps of electron beam lithography with subsequent dry etching. The nanophotonic devices were prepared from high quality silicon carrier wafers, containing 200 nm of stoichiometric silicon nitride deposited by low pressure chemical vapour deposition (LPCVD) on top of 2000 nm buried oxide. Prior to lithography, all samples were cleaned via sonication in acetone for 10 min, rinsing in isopropanol followed by oxygen plasma etching (Diener Femto, 20 % power, 10 sccm, 0.4 mbar, 3 min) to remove organic residue. To drive off remaining water the samples were dried on a hot plate at 200 °C for 5 min. During the first lithography step the pattern for electrical contacts was written in 250 nm of PMMA on a Raith EBL system. After exposure the sample was developed in a solution of methylisobutylketon (MIBK) : isopropanol (1:3). For manufacturing of metal contacts 50 nm gold was evaporated with a 5 nm adhesive layer of chromium and a 10 nm coating layer of chromium on top. The subsequent liftoff was performed by immersing the sample in acetone with subsequent weak sonication. In a second lithography step nanophotonic waveguides were defined in ZEP 520A positive resist with alignment accuracy better than 20 nm. After developing for 50 s in Xylene, reactive ion etching was used to transfer the pattern from resist into the silicon nitride on an Oxford 80 system. The etching recipe contained 50 sccm $CHF_3$ and 2 sccm $O_2$ at 175 W RF power and a base pressure of 55 mTorr. The sample was etched for roughly 90 s at a rate of 1.1 nm/s.

Carbon nanotube dispersions were prepared using the sorting method previously described by Flavel et al. [33]. In brief 10 mg of raw single-walled carbon nanotube material (SWNT) from NanoIntegris was suspended in 15 mL of $H_2O$ with 1 wt-% sodium dodecyl sulfate (SDS) using a tip sonicator (Bandelin, 200 W maximum power, 20 kHz, in pulsed mode with 100 ms pulses) applied for 2 h at ~20 % power. The resulting dispersion was then centrifuged at ~100,000 g for 1.5 h and carefully decanted from the pellet that was formed during centrifugation. The centrifuged SWNT material was then used for gel filtration fractionation. Gel filtration was performed as described previously by Moshammer et al. [34], using a Sephacryl S-200 gel filtration medium in a glass column of 20 cm length and 2 cm diameter with a final bed height of ~14 cm. After separation of the metallic SWNTs from the semiconducting SWNTs (stuck on the gel) the pH of the 1 wt-% SDS in $H_2O$ eluent was changed from 4 to 1 upon addition of the appropriate concentration of HCl. The pH was reduced in 12 steps with the (8,7) material collected as an early fraction. The collected fraction was then dialyzed for 24 h to readjust the pH to 7 in 1 mL Float-A-Lyzer G2 dialysis devices by using 500 mL of a 1 wt-% sodium cholate solution in $H_2O$. The absorption spectrum of the dispersion is shown in the supplementary information Fig. S2.

Carbon nanotubes were deposited onto the electrodes by AC-dielectrophoresis. The CNT-suspension was diluted 1:100 in deionized water and a 20 µl droplet of the aqueous suspension was placed on the chip. The electric field (2 $V_{p-p}$ AC voltage at 10 MHz) was applied to the common electrode with a function generator SRS DS345. After 5 min the sample was rinsed with water and methanol and the field was removed. Samples were annealed at 150° C for 2 hours in oven to improve the contact adhesion. For further details we refer to [25,26,35] Devices were measured at room temperature in air without further treatments.

**Characterization of waveguides and couplers with an external light source**

Light from a supercontinuum white light source (model number Leukos-SM-30-UV) was coupled into an optical fiber array (supplementary information Fig. S3a-b). This array consists of eight fibers with a core size 8.5 µm and a distance between two adjacent fibers of 250 µm. The on-chip grating couplers were designed in accordance to this distance (250 µm separation for the long waveguides and 500 µm for the Mach-Zehnder interferometer devices). The end facet of the fibers are polished at an angle of 8° to reduce back reflections as commonly done in angle polished fiber connectors. Light from one fiber is coupled into the chip plane through Bragg diffraction on a grating coupler. After propagation through the device the transmitted signal is coupled out on a second grating coupler port and measured with a spectrometer (JAZ Spectrometer System, Ocean Optics) to acquire a spectrum in a range of 340-1014 nm with a spectral resolution of 0.3 nm. To minimize coupling losses the fiber array has to be in close proximity to the sample surface. Therefore the fiber array was mounted on a movable stage, which could be precisely controlled with a piezo actuator (Picomotor, New Focus). For coarse alignment of the sample to the fiber array an optical microscope with sufficient focal length was employed. For a fine alignment additional piezo stages were used to optimize the transmission signal, which was simultaneously recorded with a photoreceiver.

**Finite-difference time-domain calculation of the electric-field intensity**

In order to analyze the near-field coupling of the CNT emitter to the waveguide structures FDTD simulations were performed using the commercial software package OmniSim distributed by Photon Design. The simulations were carried out full-vectorial in three dimensions using high grid resolution to ensure convergence within a 5% error margin. The CNT emitter was modeled as an elongated dipole source with dimensions corresponding to the real nanotube emitters, placed in direct contact with the top surface of the waveguide.

Simulations were carried out both for continuous-wave excitation to obtain travelling wave modal distributions as shown in Fig.1d, as well as with pulsed excitation to obtain the broadband response of the waveguide structure.

**Electroluminescence measurement setup**

Samples were mounted underneath a Zeiss AxioTech Vario microscope, directly attached to an Acton Research SpectraPro 2150i spectrometer and a Princeton Instruments PIXIS 256E Silicon CCD camera (1024x256 pixels, -60°C), all within a light-tight box (supplementary information Fig. S3d-e. The spectrometer can operate in the imaging mode, with a mirror to take real-space images, or in the spectroscopy mode, with a diffraction grating (300 grooves/mm, 750 nm blaze wavelength). The samples were mounted on a motorized stage, electrically contacted with probe needles and biased with a Keithley 6430 SourceMeter. Electroluminescence images and images under external illumination were recorded with Zeiss EC Epiplan-Neofluar 20x/0,50 and Motic Plan Apo 20x/0.42 objectives. Higher-resolution images were recorded with Zeiss LD EC Epiplan-Neofluar 100x/0,75 and Mitutoyo M Plan Apo 100x/0.70 objectives. The dark current of the CCD detector yields about 2 counts per pixel per hour, and allows integrating the signal over extended periods. We estimate the sensitivity of the setup on the basis of the quantum efficiency and amplifier gain of the CCD, the efficiency of the grating, the geometrical constraints of the microscope optics (optical path), and by assuming an isotropic emitter, to about 100 emitted photons per count, in reflection and diffraction mode, respectively, and the bandwidth per pixel $\Delta\lambda$ is 0.5 nm. The wavelength axis was calibrated with a mercury lamp, and the relative spectral response of the system has been measured with a calibrated halogen lamp. All recorded spectra were corrected accordingly. The spatial resolution is 0.26 µm and 1.34 µm for 100x and 20x objectives, and the spectral resolution is 1.5nm.

**Acknowledgements**

W.H.P. acknowledges support by DFG grant PE 1832/1-1 and PE 1832/1-2. B.S.F. acknowledges support by the Humboldt Foundation and support from the DFG under grant FL834/1-1. We also acknowledge support by the Deutsche Forschungsgemeinschaft (DFG) and the State of Baden-Württemberg through the DFG-Center for Functional Nanostructures (CFN) within subprojects A6.04 and B1.09. The authors thank Michael Engel, Martin Wegener and Ferdinand Evers for helpful discussions and Silvia Diewald for technical assistance.

Supplementary information accompanies this paper on the grating coupler design, absorption spectrum of CNT dispersion, experimental schemes, fitting procedure on electroluminescence spectra and the group refractive index of $Si_3N_4$ waveguides.

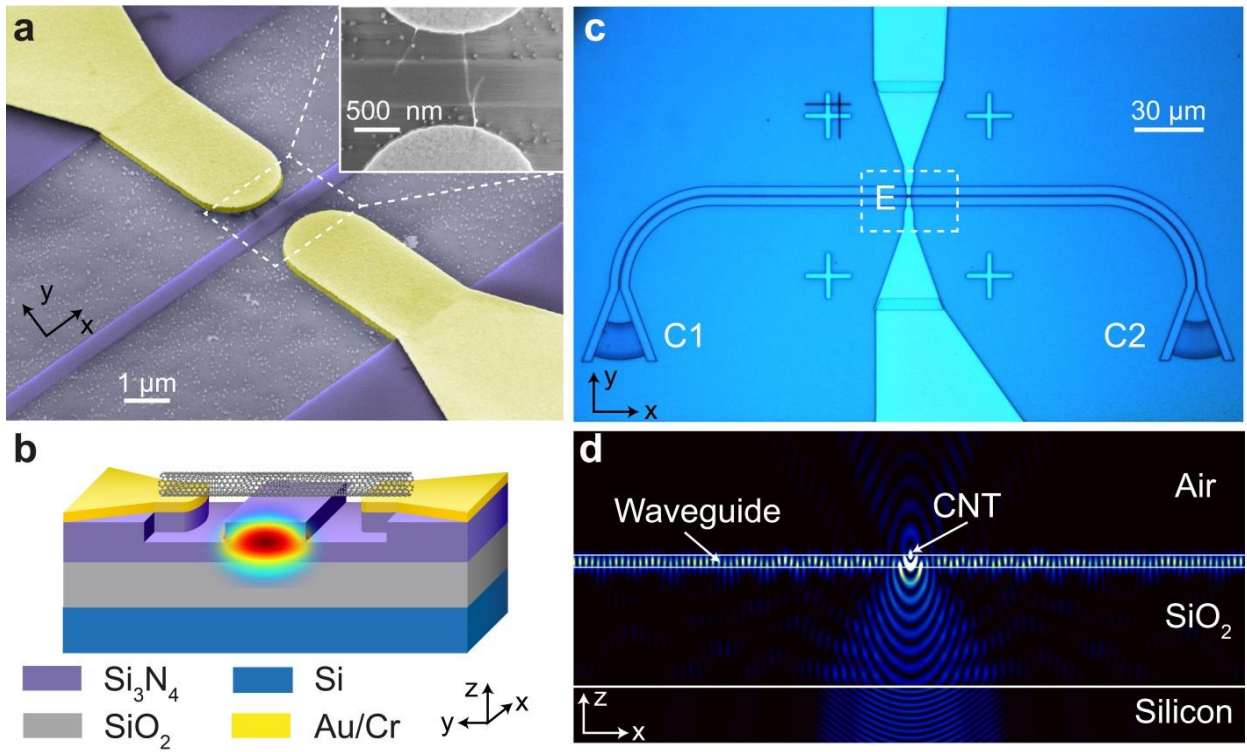

**Figure 1 | Waveguide-coupled carbon nanotube light emitter. a)** False-colored scanning electron micrograph showing two metal electrodes (yellow) and a photonic waveguide (purple), bridged by a single-walled carbon nanotube. Tilt angle 45°, scale bar 1 µm. The inset shows the indicated region at higher magnification. Carbon nanotubes appear as thin white lines. Tilt angle 0°, scale bar 500 nm. **b)** Schematic cross-sectional view of the multi-layer device structure (not to scale). The central waveguide is etched into the $Si_3N_4$ layer and runs along the x-axis. The carbon nanotube is in contact with the Au/Cr metal and the waveguide, and aligned with the y-axis. **c)** Low-magnification optical micrograph of a complete device comprising the indicated central emitter region, shown in a), and the extended photonic waveguide with two terminating coupler gratings C1 and C2. Top view, scale bar 30 µm. **d)** Cross-section through a 3D finite-difference time-domain simulation of the electric-field intensity $\vec{E}^2$. The carbon nanotube is simulated as an electrical dipole at the waveguide-air interface, oscillating perpendicular to the x-z-plane.

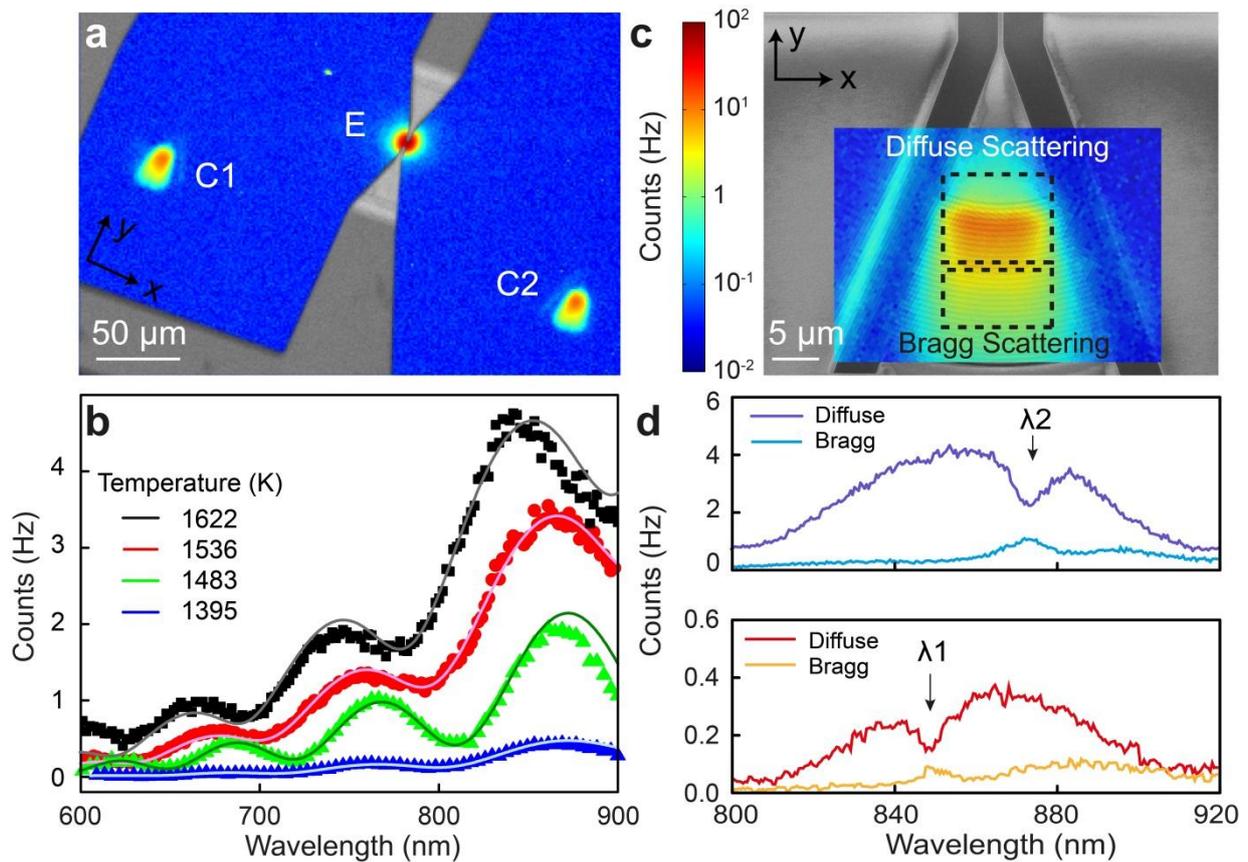

**Figure 2 | Light emission and propagation. a)** CCD-camera image of the device shown in Fig. 1a-b under electrical bias. Light emission is observed from the carbon nanotube emitter (E) and from the terminating coupler gratings C1 and C2, both connected with (E) through the waveguide (not visible). Superimposed is a grayscale image of the sample under external illumination to reveal the position of the electrodes. Scale bar 50 µm. **b)** Sequence of carbon nanotube emission spectra recorded at (E) with increasing electrical power dissipation. The data is fitted with a Planck spectrum modulated by substrate induced interference fringes. The fit-parameter temperature is given for every curve. **c)** High-resolution CCD camera image of the coupler grating C2 (with 85% opacity). Indicated are regions of diffusive scattering and Bragg scattering, close and remote from the waveguide entrance, respectively. Superimposed is a scanning electron micrograph of C2 to reveal the position of the waveguide (entering from top). Scale bar 5 µm **d)** Spectra from C1, recorded at regions of diffusive scattering and Bragg scattering, respectively. Spectra from diffusive scattering are similar to the spectra of (E) although with dips at $\lambda_1$ and $\lambda_2$. The missing intensity is recovered in the spectra from the Bragg scattering as explained in the text.

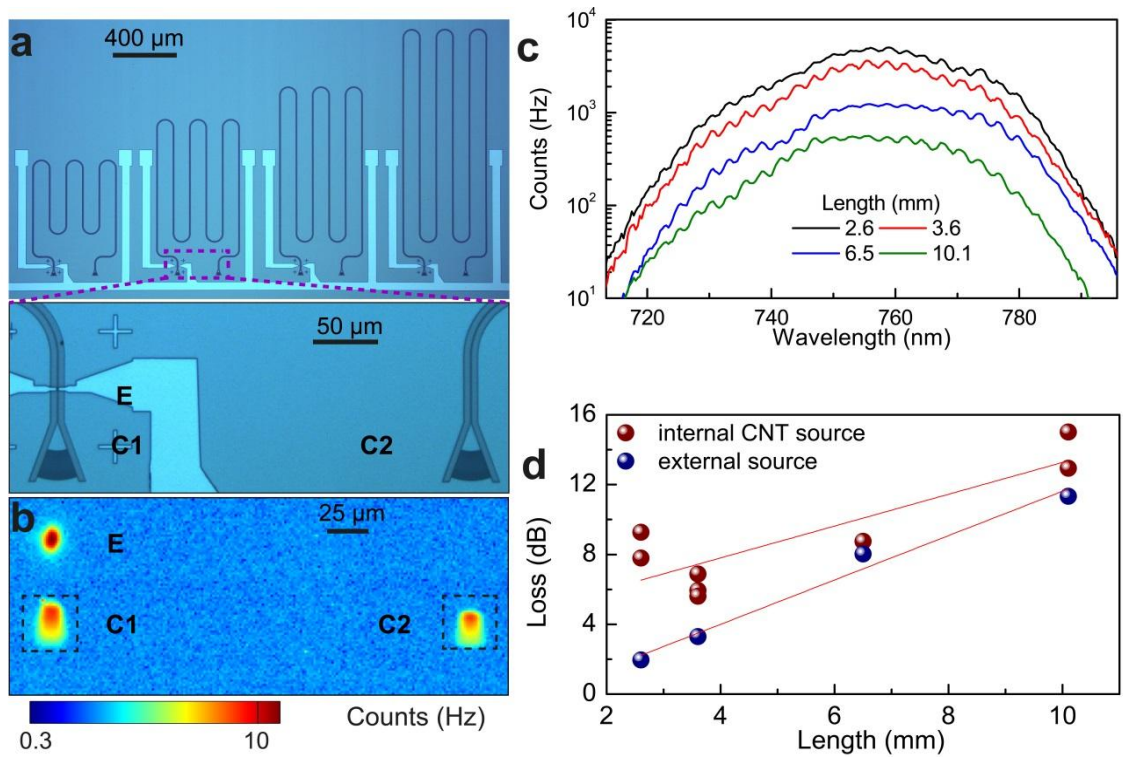

**Figure 3 | Propagation loss in extended waveguide structures. a)** Optical micrographs of 4 out of 7 devices with asymmetrically long waveguide segments between the nanotube emitter E and the coupler gratings C1 and C2. The segment between E and C1 is fixed to 25 µm. The total waveguide length L between C1 and C2 for the devices shown here is 3.6 mm, 5.3 mm, 6.5 mm and 7.3 mm. Scale bars indicated. **b)** CCD-camera image of the device with L = 6.5 mm under electrical bias. Light emission is observed from the carbon nanotube emitter (E), the nearby coupler C1 and the remote coupler C2. Losses in the waveguides are calculated from the integrated scattering light intensities I(C1) and I(C2) at C1 und C2 within the dashed regions, respectively. Scale bar 25 µm. **c)** Transmission spectra of complete devices (C1-waveguide-C2) measured with an external supercontinuum light source. Losses in the waveguide are calculated from the decreasing signal intensity at the peak wavelength 756 nm with increasing waveguide length L and are plotted in d). **d)** Comparison between losses in the waveguide determined with the external supercontinuum light source (blue) and with the integrated waveguide-coupled carbon nanotube light emitter source (red).

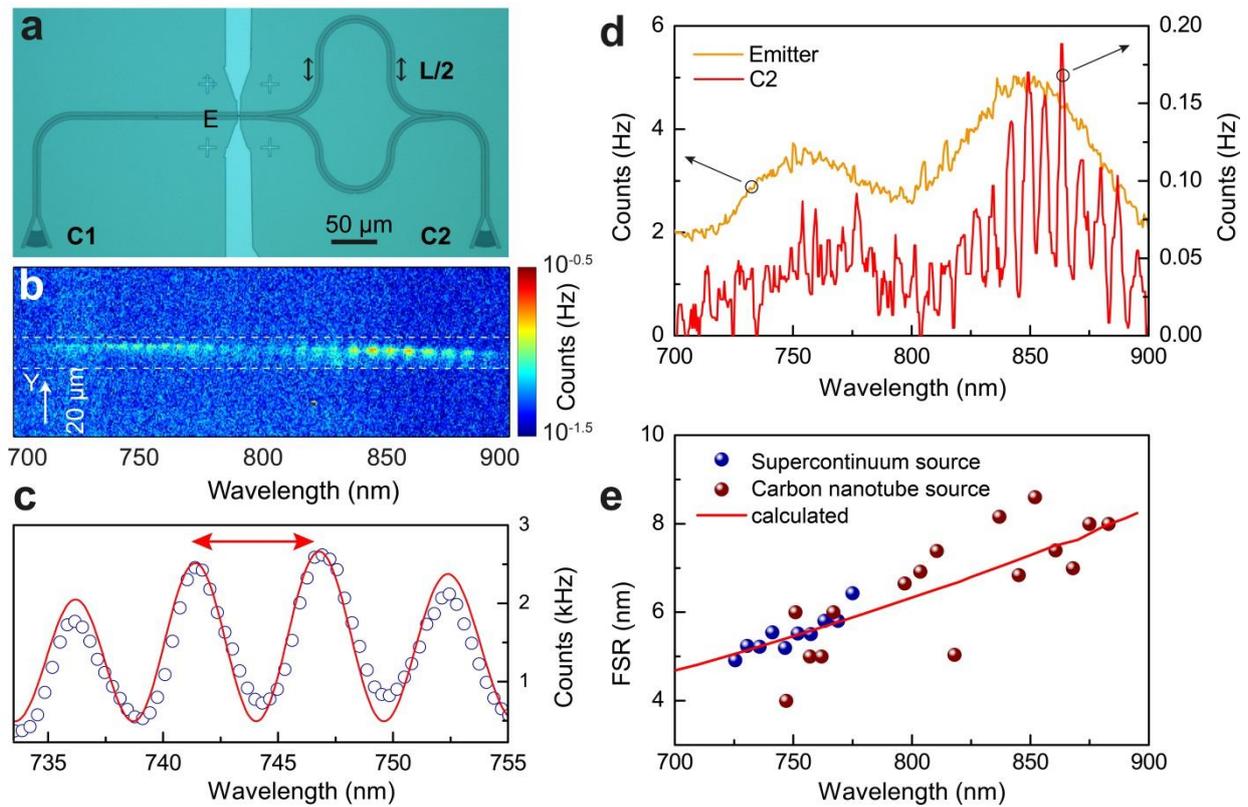

**Figure 4 | On-chip Mach-Zehnder optical interferometer. a)** Optical micrograph of a Mach-Zehnder (MZ) interferometer device. The waveguide in between the nanotube emitter E and the coupler grating C2 is split into two arms with a path difference of $\Delta L = 50$ μm. The total waveguide length between C1 and C2 is 640 μm and 640 μm + $\Delta L$, respectively. Scale bar 50 μm. **b)** CCD-camera image with combined spectral and spatial information of light emitted from the nanotube E, after passing through the MZ resonator and the coupler grating C2. The dashed lines mark the position of C2 in the y-direction and the MZ induced intensity oscillations. **c)** Transmission spectra of a complete MZ resonator device measured with an external supercontinuum light source. The red line is a fit to the data (see text). An arrow indicates a free spectral range (FSR=$\Delta\lambda$). **d)** Comparison of carbon nanotube emission spectra recorded at E (orange) with spectrum recorded at C2 after passing through the MZ resonator (red, integrated intensity from b)). **e)** Free spectral range (FSR) versus wavelength extracted from the data shown in c) and d).

# Waveguide-integrated electroluminescent carbon nanotubes
# Supplementary information

Svetlana Khasminskaya[1]*, Feliks Pyatkov[1,2,3]*, Benjamin S. Flavel[1], Wolfram H. Pernice[1,2], Ralph Krupke[1,2,3]

[1] Institute of Nanotechnology, Karlsruhe Institute of Technology, 76021 Karlsruhe, Germany

[2] DFG Center for Functional Nanostructures (CFN), 76031 Karlsruhe, Germany

[3] Department of Materials and Earth Sciences, Technische Universität Darmstadt, 64287 Darmstadt, Germany

* The authors contributed equally to this work.

Corresponding authors: R. Krupke (krupke@kit.edu), W.H. Pernice (wolfram.pernice@kit.edu)

**Grating coupler design**

To couple light into and out of the fabricated chips grating couplers as shown in Figure S1a were used [1,2]. The couplers provide phase matching between a particular waveguide mode and an unguided optical beam which is incident at $\theta = 8°$ to the surface normal in our case. As a type of Bragg grating, the basic operation of a coupler is defined by the Bragg condition [3]

$$n_{eff} = n_{top}\sin(\theta) + m\frac{\lambda}{\Lambda}$$

where $n_{eff}$ is the effective refractive index of the grating, $n_{top}$ is the refractive index of the material on top of the grating, $\theta$ is the coupling angle measured perpendicular to the chip surface, $m$ is the particular diffraction mode, $\lambda$ is the wavelength of incident light, and $\Lambda$ is the grating period. This equation describes the modes of operation for a grating at a given coupling angle, providing a Gaussian coupling profile centered around a peak wavelength as shown in Figure S1b for four different devices.

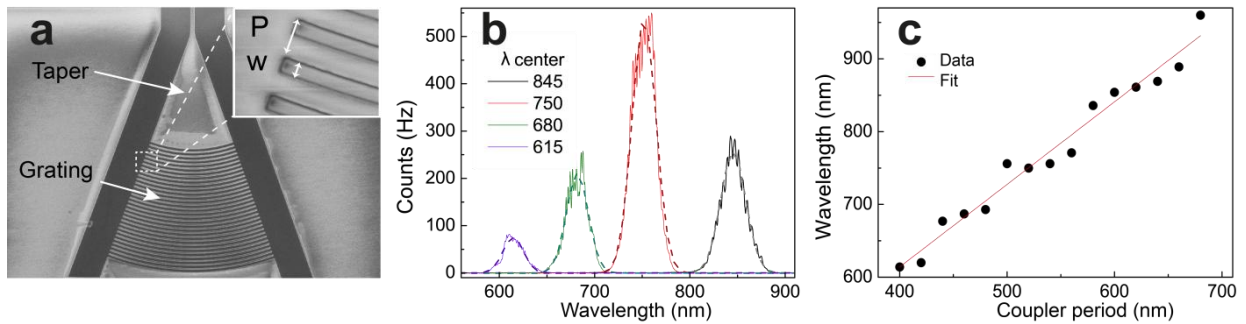

**Figure S1 | a)** SEM micrograph of a grating coupler. Inset: a zoom into the grating section with assignment of geometric dimensions. **b)** A Gaussian fit (dashed lines) to the measured data (solid lines) reveals the central coupling wavelength for devices with different grating periods. **c)** The linear dependency of the fitted coupling wavelength on the grating period allows for designing grating couplers for desired wavelengths.

From the above relation it is evident that the central coupling wavelength varies linearly with the grating period which is also found experimentally as shown in Figure S1c. Therefore a desired coupling wavelength can be conveniently chosen by adjusting the grating period.

**Absorption spectrum of CNT dispersion**

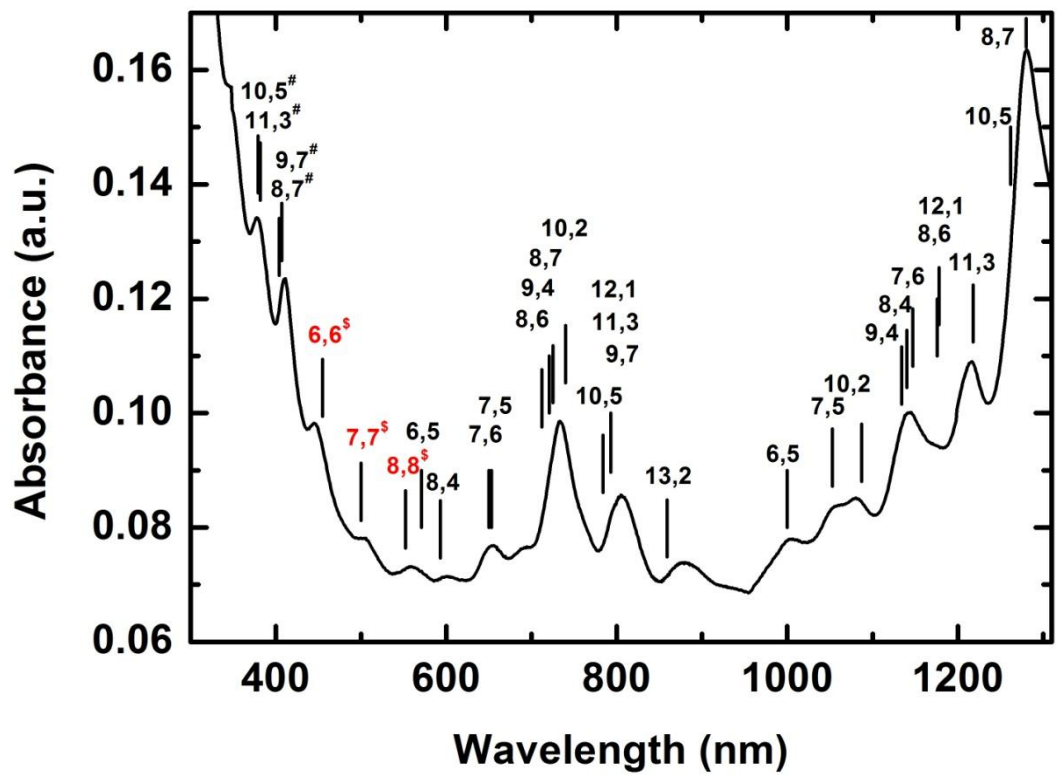

**Fig. S2 |** Absorption spectrum of the CNT dispersion. The (n,m) assignment has been obtained on the basis of photoluminescence spectra (not shown), and the data from Haroz et al.[4] and Haroz et al.[5] marked by ($) and (#), respectively. Metallic CNTs are highlighted in red.

**Measurement setups and configuration**

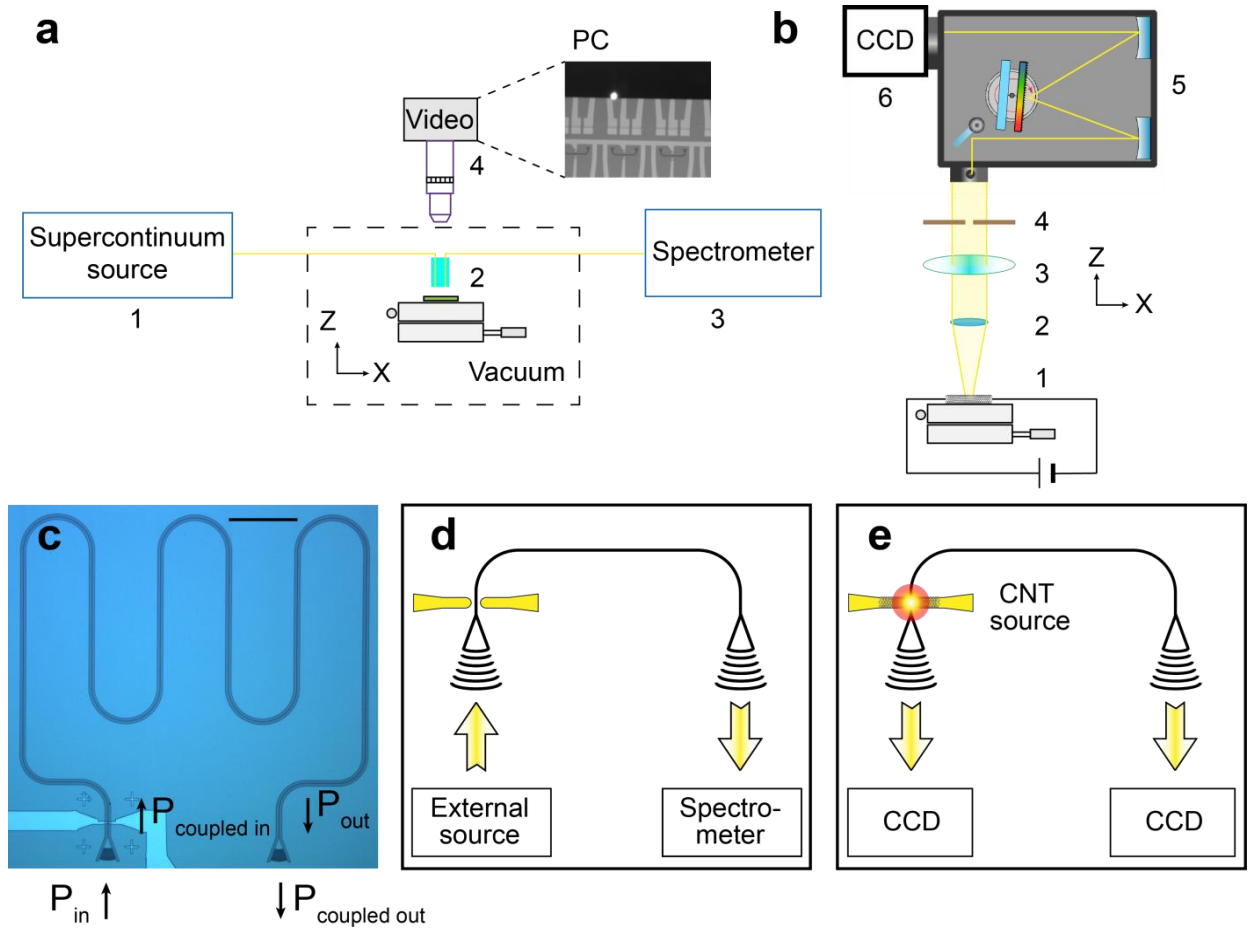

**Figure S3 |** Schematic of the two measurement setups used in the experiments. **a**) Characterization of the samples using a fiber-coupled external supercontinuum light source (1) and a fiber-coupled spectrometer (3). The sample is mounted on a piezo stage and aligned to a fiber array (2) using a CCD camera (4). **b)** Free-space characterization setup. The sample is mounted on a piezo statge (1), imaging and spectroscopy of the emitted light occurs via a microscope (2-4) coupled to a high-sensitivity CCD camera (6) / spectrometer (5). **c)** Optical micrograph of a 1.6 mm long waveguide. Scale bar 100 µm. **d)** Measurement configuration for characterization of samples using external illumination. **e)** Measurement configuration for characterization of samples under internal illumination.

**Electroluminescence spectra - fitting procedure**

All spectra acquired directly from the CNT emitter as well as from a grating show clear interference patterns with peaks around 670 nm, 740 nm, 800 nm, 850 nm, 900 nm and 950 nm. The exact spectral position of the maxima depends on the particular substrate structure. In our case, the spectra, measured at the detector, show the typical profile of black-body radiation enveloped with the interference fringes due to back-reflection from the underlying silicon substrate. The overall increase of the signal intensity at longer wavelengths is hence due to the high-energy dependence of the Planck function and thus depends on the carbon nanotube emitter temperature.

We have fitted the recorded spectra to $f(\lambda) \propto f_{Planck}(\lambda, T) \cdot [1 + 2 \cdot \gamma_{12}(\tau) \cdot f_{int}(\lambda)]$, with the temperature $T$ dependent Planck curve $f_{Planck}(\lambda, T)$, the substrate dependent interference term $f_{int}(\lambda)$, and the complex degree of coherence function $\gamma_{12}(\tau)$. The CNT source emits light directly from the air/$Si_3N_4$ interface into the microscope objective and indirectly via reflections at the $Si_3N_4$/$SiO_2$ and the $SiO_2$/Si interfaces, with the beams interfering with each other in the focal plane of the objective. Therefore the interference function calculated with a transfer matrix method was modeled as $f_{int}(\lambda)=\text{Re}(r(\lambda) \cdot r^*(\lambda))$, where $r(\lambda)=1-k \cdot reflectance(\lambda)$ and $reflectance(\lambda)$ is the reflected Fabry-Perot interference term. The coefficient $k$ represents the fraction of light that is streamed into the substrate.

To obtain the best fitting parameters, i.e. temperature, silicon nitride thickness, coefficient $k$ and degree of coherence, the sum of residuals (squares of offsets) was minimized numerically. For realistic fitting values appropriate boundary limits for each fit parameter were used in order to avoid numerical walk-off. Best fits to the measured data have been obtained with 85 nm $Si_3N_4$ and 2000 nm $SiO_2$. These values are in good agreement with the nominal SiN layer thickness after etching (90±15 nm) and the thickness of the $SiO_2$ (1960nm ±25 nm), which were obtained from independent white-light interference measurements (Filmetrics F20e-UV). Fitting values for $T$ and $\gamma_{12}(\tau)$ are given in the main text.

**Group refractive index of Si$_3$N$_4$ waveguides**

The group refractive index $n_g(\lambda)$ of the waveguide structure has been calculated according to the following equation

$$n_g(\lambda) = n_{eff}(\lambda) - \lambda \cdot \frac{dn_{eff}(\lambda)}{d\lambda},$$

where $n_{eff}(\lambda)$ is the effective refractive index of the waveguide [3]. $n_{eff}(\lambda)$ has been calculated using finite element simulations (Comsol Multiphysics). The modeled geometry comprises a 500 nm wide rib waveguide which is partially etched 100 nm into a 200nm thick Si$_3$N$_4$ layer on top of 2000 nm SiO$_2$. These are the nominal waveguide dimensions used in the experiment, as described in the main text.

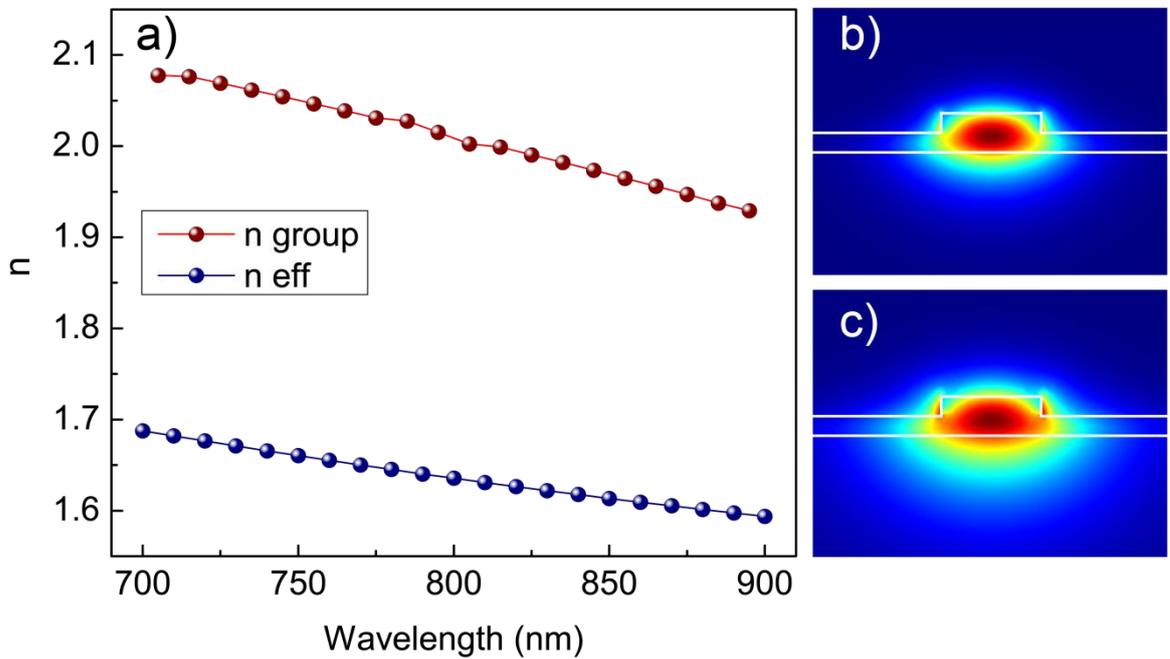

**Figure S4 | a)** Numerical simulation of the group index $n_g(\lambda)$ (red markers) and effective refractive index $n_{eff}(\lambda)$ (blue markers) of the waveguide. **b)** Cross-sectional electric field intensity distribution $\vec{E}^2$ for $\lambda=700$ nm and $\lambda=900$ nm (**c)**), equivalent to the color overlay shown in the main text in Fig. 1b) for $\lambda=750$ nm.

The wavelength dependence of the chromatic dispersion in silicon nitride has been accounted for through a Sellmeier equation [6].